\documentclass[preprint,
aps,amsmath,byrevtex,prd,titlepage,
nofootinbib]{revtex4-2}

\usepackage{amsmath}
\usepackage{amssymb}
\usepackage{bm}
\usepackage{hyperref}
\usepackage{graphicx}
\usepackage{slashed}
\usepackage{dcolumn}
\usepackage{hyperref}

\newcommand{\beq}{\begin{equation}}
\newcommand{\eeq}{\end{equation}}
\newcommand{\eq}[1]{Eq.~(\ref{#1})}

\begin{document}

\title {Muonium Hyperfine Splitting Uncertainty Revisited}
\author {Michael I. Eides}
\email[Email address: ]{meides@g.uky.edu}
\affiliation{Department of Physics and Astronomy,
University of Kentucky, Lexington, KY 40506, USA}

\begin{abstract}
Uncertainty of the theoretical prediction for the hyperfine splitting in the ground state of muonium is considered. It is compared with the respective discussion in the two most recent CODATA adjustments of the fundamental physical constants. 
\end{abstract}

\maketitle




High precision measurements of hyperfine splitting (HFS) in muonium  for a long time were considered as a test of the high precision quantum electrodynamics (QED) and a source for precise values of the fine structure constant $\alpha$ and the muon-electron mass ratio $m_\mu/m_e$ \cite{Eides:2000xc,Eides:2007exa,Mohr:2000ie}. While the role of muonium HFS in determining the fine structure constant was made obsolete by the highly precise $\alpha$ obtained from the measurements of the electron anomalous magnetic moment $a_e$ \cite{Fan:2022eto} and the recoil frequency of the $^{133}$Cs \cite{Parker:2018vye} and $^{87}$Rb \cite{Morel:2020dww} atoms, it remains the best source for the precise value of the muon-electron mass ratio. The hyperfine splitting in muonium and muon-electron mass ratio were extracted from the Zeeman splitting measurements some time ago in two LAMPF (Los Alamos Meson Physics Facility) experiments \cite{Mariam:1982bq,Liu:1999iz} 

\beq \label{exphfslap}
\Delta \nu^{ex}_{\scriptscriptstyle HFS}(\mbox{Mu})=4~463~302~776~(51)~\mbox{Hz}, \quad \delta=1.1\times 10^{-8},
\eeq

\beq \label{massratioexp}
\left(\frac{m_\mu}{m_e}\right)_{ex}=206.768~277~(24),\quad \delta=1.2\times10^{-7}.
\eeq

\noindent
The theoretical QED formula for HFS in muonium has the form \cite{Eides:2000xc,Eides:2007exa,Mohr:2000ie}

\beq \label{formtheqed}
\Delta \nu_{\scriptscriptstyle HFS}=\nu_{\scriptscriptstyle F}\left[1+F\left(\alpha,Z\alpha,\frac{m_e}{m_\mu}\right)\right]+\Delta\nu_{\scriptscriptstyle weak}+\Delta\nu_{\scriptscriptstyle hadr}+\Delta\nu_{\scriptscriptstyle th},
\eeq

\noindent
where $\nu_{\scriptscriptstyle F}$ is the Fermi frequency 

\beq \label{fermifr}
\nu_{\scriptscriptstyle F}=\frac{16}{3}Z^4\alpha^2 \frac{m_e}{m_\mu}
\left(\frac{m_r}{m_e}\right)^{3}c\:R_{\infty},
\eeq

\noindent
$\Delta\nu_{\scriptscriptstyle weak}$ is the $Z$-boson exchange contribution \cite{Eides:1995sq,Kinoshita:1995mt,jrs1996,Asaka:2018qfg}, $\Delta\nu_{\scriptscriptstyle hadr}$ is the hadron vacuum polarization contribution \cite{Nomura:2012sb,Shelyuto:2018ejm,Keshavarzi:2019abf,Karshenboim:2021jsc} and $\Delta\nu_{th}$ is the estimate of the yet uncalculated terms, see, e.g., \cite{Eides:2000xc,Eides:2007exa}. Numerically \cite{Eides:2018rph}\footnote{This is an updated value of HFS, which takes into account some minor changes in theoretical contributions and values of the fundamental constants, which happened  after  \cite{Eides:2018rph} was published, see \cite{Eides:2018rph,Karshenboim:2021jsc,Eides:2021wuv,Adkins:2022coe,Keshavarzi:2019abf,Eides:2023ltp}. Only the last digit changed by one, what  does not change any conclusions of \cite{Eides:2018rph}.
}

\beq \label{theorpredmhfs}
\Delta \nu^{th}_{\scriptscriptstyle HFS}(\mbox{Mu})=4~463~302~873~(511)~(70)~(2)~\mbox{Hz},
\eeq

\noindent
where the first uncertainty is due to the uncertainty of $(m_\mu/m_e)_{ex}$, the second one is due to the uncalculated theoretical terms, and third is due to the uncertainty of $\alpha$. This last uncertainty is too small for any practical purposes and can be safely omitted. Combining uncertainties we obtain

\beq \label{theprhfsper}
\Delta \nu^{th}_{\scriptscriptstyle HFS}(\mbox{Mu})=4~463~302~873~(515)~\mbox{Hz},\quad \delta=1.2\times 10^{-7}.
\eeq

\noindent
The dominant contribution to the uncertainty of this theoretical prediction comes from the experimental uncertainty of the mass ratio $(m_\mu/m_e)_{ex}$ in \eq{massratioexp}, not from the estimate $\Delta\nu_{\scriptscriptstyle th}$ of the  still uncalculated terms in the theoretical expression in \eq{formtheqed}. This dominance is due to the expression for the Fermi frequency in \eq{fermifr}.

It should be mentioned that the estimate of the theoretical error in \cite{Mohr:2000ie,Mohr:2005zz,Mohr:2008fa,Mohr:2012tt,Mohr:2015ccw} is roughly two times lower than in \eq{theprhfsper}. This discrepancy on the value of the uncertainty of theoretical QED prediction for HFS in muonium exists in the literature for more than twenty years, see e.g., \cite{Mohr:2000ie,Eides:2000xc}. The reasons for this disagreement were discussed in \cite{Eides:2018rph}, but despite the extended discussions the authors of  \cite{Mohr:2000ie} and \cite{Eides:2000xc,Eides:2018rph} failed to reconcile their points of view. We will  not return to this disagreement below, an interested reader can consult the papers cited above. What is crucial for the discussion below, the authors of \cite{Mohr:2000ie,Eides:2000xc,Eides:2018rph} agree on the principal point, namely, that the theoretical value of HFS should be calculated by using the QED theoretical expression and the best available values of the physical constants \cite{Mohr:2000ie,Eides:2018rph}. 

Our goal here is to discuss the treatment of the theoretical QED prediction and its uncertainty in the two latest CODATA adjustments of the fundamental  physical constants \cite{Tiesinga:2021myr,Mohr:2024kco}. Proper treatment of the theoretical QED prediction for muonium HFS is especially pertinent at this time because after a twenty years lull a new MuSEUM experiment on measuring the muonium HFS and muon-electron mass ratio is now in progress at J-PARC \cite{MuSEUM:2025cmo}.

In a sharp departure from all the earlier CODATA papers \cite{Mohr:2000ie,Mohr:2005zz,Mohr:2008fa,Mohr:2012tt,Mohr:2015ccw} the two latest CODATA adjustments \cite{Tiesinga:2021myr,Mohr:2024kco}\footnote{Discussion of muonium HFS is identical in  both these last references.} cite not the theoretical QED prediction for muonium HFS, but ``the recommended value for the muonium hyperfine splitting"

\beq \label{tesings}
\Delta\nu_{\scriptscriptstyle Mu}(th)+\delta_{\scriptscriptstyle th}(\mbox{Mu}) =4~463~302~776~(51)~ {\rm Hz}~~~~ [1.1 \times 10^{-8}]
\eeq

\noindent
This ``recommended value for the muonium hyperfine splitting" is never defined in the text but its value and uncertainty precisely coincide with the value and uncertainty of the experimental HFS in \eq{exphfslap}. This is not a surprise, if we assume that the experimental HFS was treated in \cite{Tiesinga:2021myr,Mohr:2024kco} as a fundamental constant and was included in the least square adjustment.  As we already mentioned, the last precise experiment on muonium HFS was done in 1999 \cite{Mariam:1982bq,Liu:1999iz}, and up to this moment there are no comparable in accuracy sources for the experimental value of the muonium HFS. This means that no adjustment can significantly change the values and uncertainties of muonium HFS. Therefore, the very idea of including the experimental HFS in the least square adjustment seems to be dubious. 

But this is not how ``the recommended value for the muonium hyperfine splitting" is treated in \cite{Tiesinga:2021myr,Mohr:2024kco}. By appearance, ``the recommended value for the muonium hyperfine splitting" \cite{Tiesinga:2021myr,Mohr:2024kco} in \eq{tesings} looks like a theoretical QED prediction, it is called $\Delta E_{\scriptscriptstyle \mbox{Mu}}(\mbox{th}) +\delta_{\scriptscriptstyle th}(\mbox{Mu})$. Moreover, ``the recommended value for the muonium hyperfine splitting" is explicitly interpreted in \cite{Tiesinga:2021myr,Mohr:2024kco} 
as a theoretical QED prediction, it is compared with the theoretical QED prediction in \cite{Eides:2018rph}. The authors of \cite{Tiesinga:2021myr,Mohr:2024kco} write: ``Eides (2019) gave an alternative prediction for the uncertainty of the recommended muonium hyperfine splitting." This statement demonstrates that ``the recommended muonium hyperfine splitting" is considered in \cite{Tiesinga:2021myr,Mohr:2024kco} as a theoretical QED prediction for muonium HFS since this is what was discussed in \cite{Eides:2018rph}.

We see that taken at face value the discussion in \cite{Tiesinga:2021myr,Mohr:2024kco} leads to the conclusion that ``the recommended muonium hyperfine splitting" is the theoretical QED prediction for muonium HFS. There are too many reasons why this is not so. The estimate of the theoretical uncertainty $\delta_{\scriptscriptstyle th}(\mbox{Mu})=-4(83)~ \mbox{Hz}$ in \eq{tesings} falls well inside the 85~Hz estimate of the yet uncalculated terms in this very  text \cite{Tiesinga:2021myr,Mohr:2024kco} (or $70$ Hz as estimated in \cite{Eides:2018rph}\footnote{The estimate of the yet uncalculated terms is based on the theoretical experience and can legitimately differ from one theorist to another. Anyway this minor discrepancy is irrelevant for the discussion in this paper.}). It is also hard to interpret how this $\delta_{\scriptscriptstyle th}(\mbox{Mu})$ is compatible the $51$ Hz uncertainty on the right hand side in \eq{tesings}. In its turn this uncertainty on the right hand side in \eq{tesings} is about 5 times lower than the uncertainly of the theoretical prediction in \cite{Mohr:2015ccw}, and 10 times lower that the theoretical uncertainty in \cite{Eides:2018rph}. 

Misinterpretation of ``the recommended muonium hyperfine splitting" and its uncertainty as a theoretical QED prediction for muonium HFS could lead to serious problems, especially in connection with the forthcoming results of the MuSEUM experiment \cite{MuSEUM:2025cmo}. The goal of the experiment is to reduce the experimental uncertainties of the muonium HFS and muon-electron mass ratio in \eq{exphfslap} and in \eq{massratioexp} by an order of magnitude. The experimental team hopes to measure the weak interaction contribution to the shift of an atomic energy level for the first time, and obtain limits on possible New Physics contributions to muonium HFS \cite{MuSEUM:2025cmo,kshim2018}. For this last goal a proper estimate of the uncertainty of  the theoretical prediction becomes crucial and misinterpretation of ``the recommended muonium hyperfine splitting" becomes particularly troubling. If the experimental value of HFS would turn out to be outside the error bars of the CODATA ``recommended value for the muonium hyperfine splitting" in \eq{tesings} \cite{Tiesinga:2021myr,Mohr:2024kco} this could lead to a wrong conclusion about existence of  a New Physics contribution. As discussed above, this is not the case, the true theoretical uncertainty is not 51 Hz cited in \cite{Tiesinga:2021myr,Mohr:2024kco}, but 271 Hz according to \cite{Mohr:2015ccw} and 515 Hz according to \cite{Eides:2018rph}. The above discussion shows that the misleading presentation of ``the recommended muonium hyperfine splitting" in \cite{Tiesinga:2021myr,Mohr:2024kco} as the theoretical QED prediction for muonium HFS requires immediate remedy.  Hopefully, this short note will serve this goal.


This work was supported by the NSF grant PHY-2510100.

\end{document}